\documentclass[]{JHEP}
\usepackage{graphicx}
\def\eq#1{{eq.~(\ref{#1})}}

\def\Im{\mathop{\mbox{Im}}}
\def\Re{\mathop{\mbox{Re}}}

\def\eprime{$\varepsilon'/\varepsilon$}

\newcommand{\be}{\begin{equation}}
\newcommand{\ee}{\end{equation}}
\newcommand{\bea}{\begin{eqnarray}}
\newcommand{\eea}{\end{eqnarray}}

\title{Lattice matrix elements confronting the 
experimental value of \eprime\ }

\author{Stefano Bertolini and Marco Fabbrichesi\\
INFN, Sezione di Trieste and\\
Scuola Internazionale Superiore di Studi Avanzati (SISSA)\\
via Beirut 4, I-34013 Trieste, Italy.}

\abstract{A new lattice estimate of 
$K\rightarrow 2 \pi$ transitions  claims, contrary to all
other computations, that the hadronic matrix element for
the gluon penguin operator $Q_6$ has opposite sign and, in
addition, is much larger than  the vacuum saturation
approximation.
We comment under what conditions (if any)
it is possible to reconcile this lattice result with the experimental 
value of \eprime\ .  The dramatic impact of new physics in the kaon
system that seems to be required is  not easily accommodated 
within our present theoretical understanding.}


\preprint{SISSA 102/99/EP \\
August 1999}
\begin{document}

{\bf 1.} A new lattice estimate  of 
$K\rightarrow 2 \pi$ transitions using domain-wall 
fermions claims~\cite{soni}, contrary to all
other computations, that the hadronic matrix element for
the gluon penguin operator $Q_6$ has opposite sign and, in
addition, is much larger than its vacuum saturation
approximation. 
This surprising result comes about because of the contribution 
of the so-called eye-contraction diagrams.
The resulting value for the CP violating parameter \eprime\ is of
the opposite sign and almost one
order of magnitude bigger than the current experimental determination:
\be
\varepsilon'/\varepsilon = (2.1 \pm 0.46 ) \times 10^{-3}\, , \label{exp}
\ee
which is obtained by averaging over the 
KTeV~\cite{KTeV} and NA48~\cite{NA48} preliminary results 
as well as the older 1992-93 experiments
(E731~\cite{E731} and NA31~\cite{NA31}).

If this lattice result (taken at its face value)
will stand further scrutiny it  raises serious
questions on our understanding of electroweak  physics in the kaon
system.

In this letter we would like to discuss under what conditions (if any)
it is possible to reconcile a large and positive
$\langle 2\,\pi| Q_6|K\rangle$ with the experimental 
value of \eprime\ . We consider two possible scenarios:
\begin{itemize}

\item A modification of short-distance physics that 
changes both the sign and the size of the Wilson coefficient of the gluon
or electroweak penguin operators. This can in principle
be achieved by tampering
with the initial matchings of the various coefficients while
preserving the standard basis of operators. As we shall discuss,
the behavior of the renormalization group equations (RGE) for the
relevant effective lagrangian force us to rather extreme changes
in order to achieve the desired effect. 
Even though
supersymmetry provides a framework for potentially large effects,
to construct a model in which 
such large deviations from standard physics are present in $\varepsilon'$
while conspiring to be invisible everywhere else involves a contrived
set of assumptions.

\item An enlargement of the standard operator basis. 
We limit our discussion to the case of the chromomagnetic
penguin operator because it does not affect the renormalization
of the Wilson coefficients of the other dimension six operators 
(allowing us to draw model-independent conclusions) and it
has been already shown to be a promising candidate
for new-physics effects in \eprime\ . Given the current estimate
of this operator's matrix element, short-distance changes
alone---although potentially very large---are not  sufficient 
in bringing \eprime\ in the experimental ball-park. 
A final assessment requires a lattice evaluation of the
hadronic matrix element of the chromomagnetic operator. 

\end{itemize}
Both scenarios call for a dramatic impact of new physics in the kaon
system and they are not easily accommodated within our present
theoretical understanding. 
More exotic extensions of the standard operator basis and more
extreme scenarios require a detailed 
model-dependent discussion which is beyond the scope of the present
letter.~\footnote{Possible
effects of dissipation and loss of coherence in the
kaon system relevant to \eprime\ have been discussed in 
ref.~\cite{flor}.}

\bigskip\bigskip
{\bf 2.} Let us fix our notation by introducing the $\Delta S=1$ 
effective four-quark lagrangian for $m_c < \mu < m_b$
\be
{\cal L}^{\Delta S=1}_{eff} = - \frac{G_F}{\sqrt{2}} \lambda_u \Bigl\{
 (1 - \tau) \sum_{i=1,2} C_i(\mu) \bigl[ Q_i(\mu) - Q_i^c(\mu)\bigr]
+ \tau  \sum_{i=1,10} C_i(\mu) Q_i(\mu) \Bigr\} \, , \label{lag}
\ee
where $\lambda_q \equiv V_{qd}V_{qs}^*$ and $\tau =
-\lambda_t/\lambda_u$. In the discussion that follows we will assume
as the standard model (SM) reference value for the CP phase 
$\Im \lambda_t = 10^{-4}$.
The renormalization group Wilson coefficient $C_i(\mu)$ are known to the
next-to-leading order in $\alpha_s$ and
$\alpha_e$~\cite{buras1,martinelli}.

The standard basis of effective operators is
\be
\begin{array}{rcll}
Q_{1} &  = & \left( \overline{s}_{\alpha} u_{\beta}  \right)_{\rm V-A}
            \left( \overline{u}_{\beta}  d_{\alpha} \right)_{\rm V-A}\, ,
& \quad 
Q_{1}^c  =  \left( \overline{s}_{\alpha} c_{\beta}  \right)_{\rm V-A}
            \left( \overline{c}_{\beta}  d_{\alpha} \right)_{\rm V-A}
\, , \\[1ex]
Q_{2} & = & \left( \overline{s} u \right)_{\rm V-A}
            \left( \overline{u} d \right)_{\rm V-A}\, ,
& \quad
Q_{2}^c  =  \left( \overline{s} c \right)_{\rm V-A}
            \left( \overline{c} d \right)_{\rm V-A}
\, , \\[1ex]
Q_{3,5} & = & \left( \overline{s} d \right)_{\rm V-A}
   \sum_{q} \left( \overline{q} q \right)_{\rm V\mp A} 
\, , & \\[1ex]
Q_{4,6} & = & \left( \overline{s}_{\alpha} d_{\beta}  \right)_{\rm V-A}
   \sum_{q} ( \overline{q}_{\beta}  q_{\alpha} )_{\rm V\mp A}
\, , & \\[1ex]
Q_{7,9} & = & \frac{3}{2} \left( \overline{s} d \right)_{\rm V-A}
         \sum_{q} \hat{e}_q \left( \overline{q} q \right)_{\rm V\pm A}
\, , & \\[1ex]
Q_{8,10} & = & \frac{3}{2} \left( \overline{s}_{\alpha} 
                                                 d_{\beta} \right)_{\rm V-A}
     \sum_{q} \hat{e}_q ( \overline{q}_{\beta}  q_{\alpha})_{\rm V\pm A}
 \, , & 
\end{array}  
\label{Q1-10} 
\ee
where $ V\pm A$ stands for $\gamma_\mu (1 \pm \gamma_5)$ and 
 $\hat{e}_q$ is the value of the electric charge of the quark 
$q=u,d,s,c$. 

\bigskip\bigskip
{\bf 3. } Taking the hadronic
 matrix elements of the operators in \eq{Q1-10}
obtained by the lattice calculation~\cite{soni},
 and given their standard model Wilson
coefficients, \eprime\ is completely
dominated by the contribution of $Q_6$ and is therefore
 large and negative (in dramatic disagreement with
the experiment).  

In order to reconcile the lattice and the experimental
result, we must somehow compensate for this dominant contribution. 
The ratio \eprime\ is determined by the sum of isospin $I=0$ 
and $I=2$ amplitudes. Let us examine in turns possible
modifications on these
two classes of contributions.

A first possibility, in the $I=0$ amplitude, is
that the Wilson coefficient of the $Q_6$, at the scale at
which the matrix element is computed (about 2 GeV), is
changed with respect to its standard model value
in order to reproduce the experimental result. Assuming that
new physics only modifies the initial conditions  (at $m_W$) 
of the RGE, we need to know how these changes are propagated by the
RGE down to the hadronic scale. This has been discussed
for the whole operator basis in section VII of~ref.\cite{review}.

In order to keep the discussion as model independent as possible, we
parameterize all
deviations from the SM matching conditions 
$C_i(m_W)$ in terms of  the parameters
\be
r_i \equiv C_i(m_W)/C_i^{SM}(m_W) \, .
\label{ri}
\ee 
In general, a Wilson coefficient at the low scale receive
contributions from both a multiplicative and additive renormalization,
the latter arising from QCD-induced operator
mixing. In particular, the $C_6(\mu)$ is
dominated in the standard basis by the additive renormalization induced
by the mixing of $Q_6$ with the $Q_2$ operator.

\FIGURE[t]{
\includegraphics[scale=0.7]{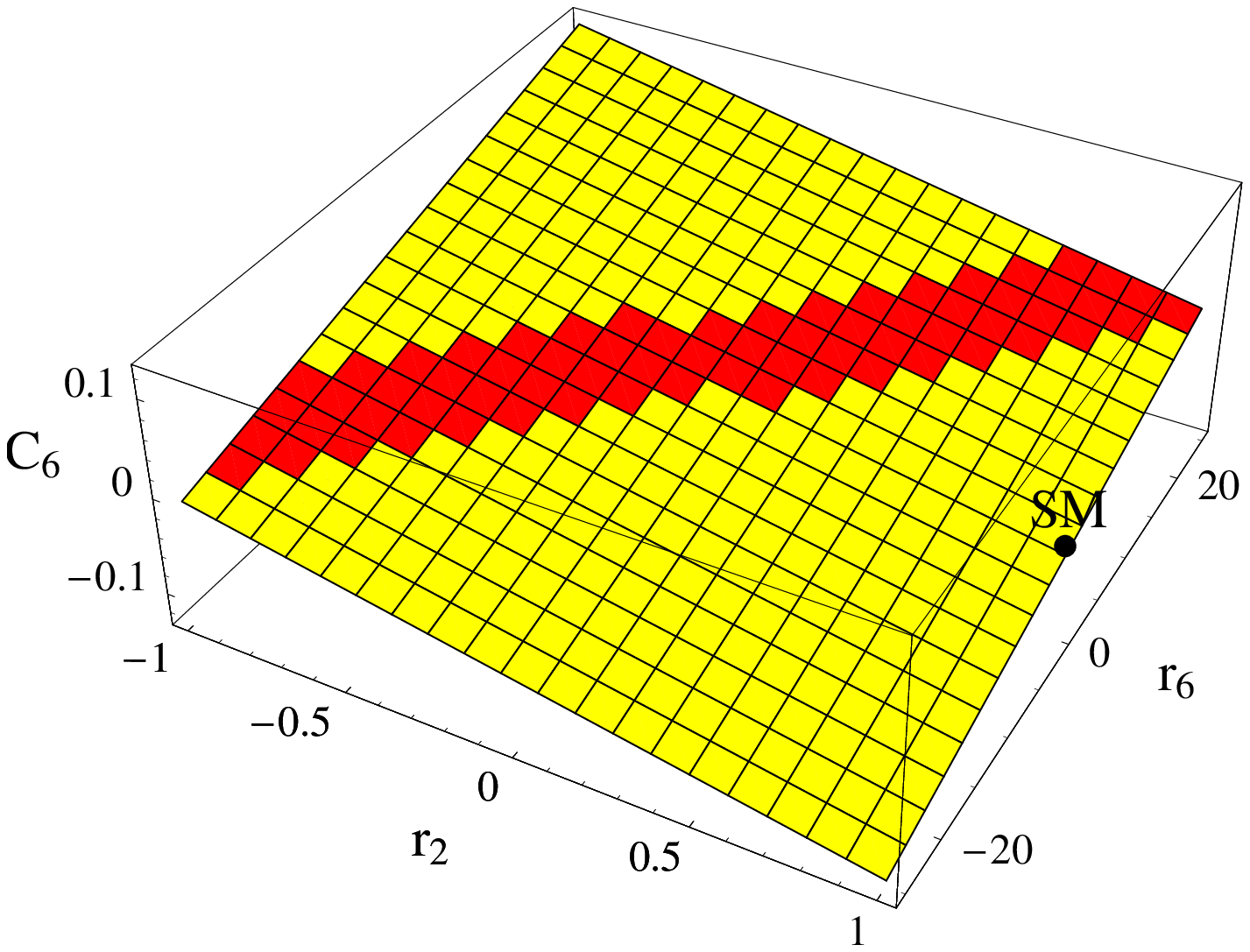}
\caption{The Wilson coefficient $C_6$(2 GeV) 
as a function of the parameters $r_2$ and $r_6$ defined in \eq{ri}.}
\label{ploty6}
}

We have plotted in Fig.~\ref{ploty6} the Wilson coefficient $C_6$(2 GeV) as a
function of the parameters $r_2$ and $r_6$.
The dark (red) band represents the values of $C_6$ 
for which the experimental result for \eprime\
 would be recovered (leaving all other
Wilson coefficients unchanged), that is
\be
- 0.2 < \frac{C_6(2\:\mbox{GeV})}{C_6^{SM}(2\:\mbox{GeV})} < - 0.1 \ . 
\label{c6}
\ee
As it can seen by inspection,
this possibility is realized either
by a drastic reduction of $C_2(m_W)$ (the left side of Fig.~\ref{ploty6})
or a large enhancement of
$C_6(m_W)$  (the right side of Fig.~\ref{ploty6}).
However, the value of $C_2$(2 GeV) is severely constrained by the 
CP-conserving amplitude $A_0(K\rightarrow 2 \pi)$, which in the same
lattice calculation is reproduced up to a factor of two. Moreover, it
is difficult to substantially change $C_2(m_W)$ by means of new
physics because this Wilson coefficient is due to tree level $W$-exchange.

Looking then at Fig.~\ref{ploty6}, and given that $C_2(m_W)$ 
cannot differ too much from its standard-model
 value, it is possible to reproduce the experimental \eprime\
only by enhancing $C_6(m_W)$ by more than a factor of twenty. 
However, such a large enhancement can hardly take place
without affecting other processes and we shall come back to
it after discussing the electroweak penguin. 

Leaving the $I=0$ contribution alone, we can still compensate for the
large and negative result  by acting on the the $I=2$
contribution. Here the dominant operator is $Q_8$. Given the
lattice estimate of the matrix element of this operator, agreement with
the experiments would require
\be
C_8(2\: \mbox{GeV}) \simeq - 30\:\: C_8^{SM} (2\: \mbox{GeV}) \, .
\label{c8np}
\ee

The RGE analysis shows that $C_8$(2 GeV) varies proportionally to
$C_7 (m_W)$ (recall that $C_8 (m_W)$=0) 
and, as one can see from Table~\ref{table1},
$C_7 (m_W)$ receives contributions from photon and $Z$ penguins.

\TABLE[t]{
\begin{tabular}{c c c c c c c c c c c}
$\mu=m_W$   &$C_1$&$C_2$&$C_3$&$C_4$&$C_5$&$C_6$&$C_7$&$C_8$&$C_9$&$C_{10}$\\
\hline
Tree             & &$\surd$& & & & & & & & \\ 
Tree + $g$       &$\surd$&$\surd$& & & & & & & & \\ 
Tree + $\gamma$  & &$\surd$& & & & & & & & \\ 
${\cal P}_g$     & & &$\surd$&$\surd$&$\surd$&$\surd$ & & & & \\ 
${\cal P}_\gamma$& & & & & & &$\surd$& &$\surd$& \\ 
${\cal P}_Z$     & & &$\surd$& & & &$\surd$& &$\surd$ & \\
${\cal B}$       & & &$\surd$& & & & & &$\surd$ & \\
\end{tabular}
\caption{Contributions to the one-loop matching of the
$\Delta S=1$ Wilson coefficients at $\mu=m_W$. 
Non-vanishing contributions to $C_8$ and $C_{10}$ arise via the
QCD renormalization of the operators $Q_7$ and $Q_9$, respectively.
The label ``Tree'' stands for tree-level $W$ exchange, ${\cal P}_{g,\gamma,Z}$
for gluon, photon and $Z$-induced penguins, 
while ${\cal B}$ for the $W$-induced box diagrams.}
\label{table1}
}

To date, the best limits on the CP conserving component of the 
$Z$-penguin operator,
$\Re{\cal P}_Z$, are provided by the $K_L\to\mu^+\mu^-$ 
decay~\cite{nir,colangelo,silvestrini}, 
whose branching ratio is measured to be~\cite{expkl}
\be
B(K_L\to\mu^+\mu^-) = (7.18\pm 0.17) \times 10^{-9}\ ,
\label{kltomupmum}
\ee
even though 
the constraint on $\Re{\cal P}_Z$ is not as accurate as the
experimental precision because of the
theoretical long-distance uncertainties related to the 
two-photon component~\cite{portoles}.

The cleanest constraint on $|{\cal P}_Z|$
comes from the decay $K^+\to\pi^+\bar\nu\nu$,
which is currently measured to be~\cite{adler}
\be
B(K^+\to\pi^+\bar\nu\nu) = 4.2\ ^{+9.7}_{-3.5}\times 10^{-10}\ . 
\label{kptopinunu}
\ee
Due to the lack of further evidence, the branching ratio
in \eq{kptopinunu} is going to be reduced by a 
factor of two or three~\cite{redlinger}. 

Taking the standard model expectation 
$B(K^+\to\pi^+\bar\nu\nu) = (0.8\pm 0.3) \times 10^{-10}$~\cite{buchalla}
as a reference point, a numerical analysis shows
that we can at most modify $\Im{\cal P}_Z^{SM}$ by a factor of
sixteen.

Barring the unlikely possibility of independent and widely
different effects
in the effective $(\bar s d)_{V-A}$ vertex
of the photon and $Z$ penguins, \eq{kptopinunu} rules out the 
enhancement required by \eq{c8np}. Analogous considerations hold for the other
electroweak operators which would entail even larger deviations
of their Wilson coefficients from the standard model values. 
By the same token, also the enhancement by a factor of twenty of the
gluonic penguin coefficient $C_6(m_W)$---which we argued is necessary
in the $I=0$ amplitude in order to reproduce the
experimental value of \eprime\ ---is difficult to accommodate.

A notable exception to the above argument is
the presence of gluino-induced flavour-changing neutral currents
in supersymmetric models. The gluino induced $\Delta S = 1$ transitions
are suppressed in Z-penguin diagrams compared to
gluon or photon penguins by a factor of $O(m_K^2/m_Z^2)$
(for a detailed discussion of
this effect which follows from gauge invariance see 
ref.~\cite{BBM}). 
As a consequence, the bounds
on Z-penguins are not effective on gluino-induced gluon and photon penguins.
However, the gluino contributions to the standard 
$\Delta S = 1$ penguin operators
are indirectly constrained by the sharp bounds on
the gluino-induced $\Delta S = 2$ box diagrams 
relevant to $\bar K^0-K^0$ transitions~\cite{box}.

Gluino-box diagrams can also play a direct role in $\Delta S = 1$
amplitudes.
In ref.~\cite{neubert}, it is shown that gluino-box diagrams
may induce a potentially large isospin-breaking contribution to the
electroweak penguin $Q_8$ while
satisfying all other bounds. However,
the large factor required by \eq{c8np}
implies a rather large mass splitting between the right-handed squark
isospin doublet ($m_{\tilde u_R}-m_{\tilde d_R} \simeq 1$ TeV)
together with a large gluino mass in order to evade the bounds from
$\Delta m_K$ and $\varepsilon$.

\bigskip\bigskip  
{\bf 4. }  An operator not usually included in the standard model
analysis of \eprime\ is the chromomagnetic penguin 
\be
Q_{11}  =  \frac{g_s}{8 \pi^2} \: \bar s \: \left[ 
m_d R + m_s L \right] 
\: \sigma_{\mu\nu}\: G^{\mu\nu} \, d \label{q11} \, , 
\ee
where $R(L) = (1 \pm \gamma_5)/2$ and $G^{\mu\nu}$ is the gluon field.

The matrix element $\langle 2 \pi | Q_{11} | K \rangle$ has been
computed in the chiral quark model and shown to arise only at $O(p^4)$ in the
chiral expansion and to be further suppressed by a $m_\pi^2/m_K^2$ 
factor with
respect to the naive expectation~\cite{noi}. For these reasons,
even though the $Q_{11}$ Wilson
coefficient receives a large additive renormalization, its standard model
contribution to \eprime\ has been shown to be negligible~\cite{noi}.

This is no longer true if the Cabibbo-Kobayashi-Maskawa (CKM)
 suppression of its CP violating
component can be lifted without violating other bounds. 
A clever example of it
 is discussed in ref.~\cite{MM}, where
the standard model factor $\Im\lambda_t \simeq O(\sin^5\theta_C)$
is replaced in a supersymmetric framework by
a CP violating squark mixing of $O(\sin\theta_C)$,
thus introducing a potential enhancement of the chromomagnetic 
Wilson coefficient by three orders of magnitude. This solution
allows for a contribution to \eprime\ at the $10^{-3}$ level
by saturating the known bounds coming from CP violating
phenomenology (discussions of the implications of these
bounds on various supersymmetric models can be found in 
ref.~\cite{susyq11}). 

This enhancement of $C_{11}$ is still not enough to compensate for the 
huge negative $Q_6$ contribution to \eprime\ of the
lattice result; in fact, keeping fixed the gluon penguin
contribution, agreement with experiment would  
require a  matrix element for  $Q_{11}$ larger by about a
factor of ten.
Actually, the leading $O(p^4)$ chiral quark model estimate of the
hadronic matrix element may receive potentially
large $O(p^6)$ contributions if the accidental
$m_\pi^2/m_K^2$ suppression is replaced by $m_K^2/\Lambda_\chi^2$.
To further assess this possibility,
it would therefore be interesting to have an estimate
of the $Q_{11}$ matrix element from the same lattice approach that
produces the large and positive matrix element for the $Q_6$.

Clearly, extensions of the standard model
 can also involve new effective operators
beyond the standard basis of \eq{Q1-10} and  $Q_{11}$. However, in this
case very little can be said without a complete re-analysis of
 \eprime\ . 
Similarly, scenarios in which the CKM matrix
 is taken to be real and CP violation arises only in the new-physics sector
 can in principle be invoked but again require a 
detailed model-dependent
analysis before being considered a viable alternative.

\bigskip\bigskip
{\bf 5.} In conclusion,
while a combination of the above scenarios may make some of
the requirements less severe, 
a viable model which avoids the
phenomenological bounds discussed is forced to rely on a contrived
choice of parameters. 
It is nevertheless remarkable how supersymmetry can provide a 
framework for potentially large effects on \eprime\ while satisfying 
all present data.

It is also fair to add that the
lattice result by means of  domain-wall fermions in ref.~\cite{soni}
must stand further scrutiny and corroboration
before concluding that the standard model is facing its most dramatic 
challenge to date.

\bigskip\bigskip
We thank F. Borzumati and M. Neubert for useful discussions and comments.

%
\renewcommand{\baselinestretch}{1}

\end{document}